\documentclass[prb,aps,twocolumn,showpacs,superscriptaddress]{revtex4}
\usepackage[dvips]{graphicx}

\usepackage{latexsym}
\usepackage{amsmath}
\usepackage{subfigure}
\usepackage{epsfig}

\begin{document}


\title{Competing orders, non-linear sigma models, and topological terms in quantum magnets}
\author{T.~Senthil}
\affiliation{Center for Condensed Matter Theory, Indian Institute of Science,
Bangalore-560012, India}
\affiliation{Department of Physics, Massachusetts Institute of
Technology, Cambridge, Massachusetts 02139}
\author{Matthew~P.~A. Fisher}
\affiliation{Kavli Institute for Theoretical
Physics University of California Santa Barbara, CA 93106-4030}


\begin{abstract}
A number of examples have demonstrated the failure of the Landau-Ginzburg-Wilson(LGW) 
paradigm in describing the competing phases
and phase transitions of two dimensional quantum magnets. In this paper we argue that
such magnets possess field theoretic descriptions in terms of their slow fluctuating orders
provided certain topological terms are included in the action. These topological terms may thus be viewed 
as what goes wrong within the  conventional LGW thinking. The field theoretic descriptions we develop are 
possible alternates to the popular gauge theories of such non-LGW behavior. Examples that are studied include
weakly coupled quasi-one dimensional spin chains, deconfined critical points in fully two dimensional magnets, and 
two component massless $QED_3$. A prominent role is played by an anisotropic 
$O(4)$ non-linear sigma model 
in three space-time dimensions with a topological theta term. Some properties of this model are discussed. 
We suggest that similar sigma model descriptions might exist for fermionic algebraic spin liquid phases.

\end{abstract}


\maketitle

Recently there has been considerable progress in
describing the competition between various kinds of ordering
tendencies of quantum spin systems in two space dimensions.
Possible Landau-forbidden direct second order quantum transitions
between Neel and valence bond solid phases of spin-$1/2$ quantum antiferromagnets 
have been described in
Refs. \onlinecite{deccp}. The corresponding critical points have
been described in terms of field theories involving gapless
bosonic spinon fields coupled to a fluctuating gauge field. They
have hence been dubbed `deconfined quantum critical points'. Other
recent work has established the stability of `algebraic spin
liquid' phases of two dimensional quantum magnets\cite{stable-u1},
at least within systematic large-$N$ expansions. These are
critical {\em phases} that exhibit power law correlations in the
spin\cite{rantwen} and other operators\cite{su4}. In the best studied cases the low energy
theory is described by a conformally invariant fixed point. All of
the existing descriptions of such algebraic spin liquids is in
terms of theories of Dirac spinon fields coupled to a fluctuating
massless gauge field. Thus these may be regarded as `deconfined
critical phases'. Despite the utility of the spinon-gauge
description neither the spinons nor the gauge photon are good
quasiparticles of the system at these deconfined critical
points/phases. Indeed it is not clear that there is {\em any}
quasiparticle description of the low energy fixed point theory.

A large number of open questions remain on such critical points/phases. Is a
description in terms of spinons and gauge fields necessary? Given
the slow power law decay for various classical order parameters,
is a description directly in terms of these orders possible? In
this paper we study a number of related questions and suggest some
interesting answers to these questions.

In contrast to two dimensions competing orders in one dimensional spin systems
are extremely well-understood: the classic example is the
one dimensional spin-$1/2$ antiferromagnetic chain. These have a
number of properties reminiscent of their two dimensional
counterparts. It is well-known that the power law phase in such
spin chains can be described by a number of different equivalent
field theories. In a semiclassical description an $O(3)$
non-linear sigma model with a topological $\theta$ term\cite{hald83} obtains
(with $\theta = \pi$). The $O(3)$ vector has the physical
interpretation of being the Neel order parameter. More useful is a
field theory in terms of an $SU(2)$ matrix $U$ with a Wess-Zumino-Witten(WZW) term\cite{1dwzw}.
The $SU(2)$ matrix $U$ again has a simple interpretation.
Indeed $U = D + i\vec N. \vec \sigma$ with $\vec N$ the Neel
vector,and $D$ the dimerization ({\em i.e} VBS or spin-Peierls)
order parameter. The VBS order parameter also has the same power
law correlations as the Neel vector in the spin-$1/2$ AF chain.
Thus the WZW field theory is written precisely in terms of the
classical order parameters that have slow correlations in this
algebraic spin liquid phase. However the action necessarily
involves the non-trivial `topological' WZW term. Finally a gauge
description of the spin-$1/2$ chain in terms of fermionic Dirac
spinons coupled to fluctuating $U(1)$ or $SU(2)$ gauge fields is also
possible\cite{hoso,mrfr}. It is readily seen using bosonization techniques that
this is exactly equivalent to the $SU(2)$ level-$1$ WZW theory.

Motivated by this we may speculate that two dimensional quantum magnets (including
possible deconfined critical
points/phases) also possess field theoretic descriptions in terms of
their slow fluctuating orders provided certain topological terms
are included in the action. Here we will collect a number of
evidences supporting this suggestion.

We first study quasi one dimensional systems of weakly
coupled one dimensional spin-$1/2$ chains. If the interchain
exchanges are unfrustrated it is expected that two dimensional
long range Neel order will develop. Frustrating interchain
exchanges promote VBS ordering of the columnar dimer pattern with
a two fold degenerate ground state. One approach to think about
the competition between these two distinct ground states is in
terms of a $2+1$ dimensional $O(3)$ non-linear sigma model with
appropriate Berry phases for the hedgehog topological defects\cite{hald88}. It is known that the
hedgehogs are doubled in a continuum description, and their
proliferation leads to VBS order in the paramagnet\cite{ReSaSuN}.
In this approach the Neel order parameter is simply represented as the $O(3)$ vector. On the other hand 
the VBS order parameter has a more complicated representation and is identified as the hedgehog topological defect.
Thus this approach treats the Neel and VBS orders on an unequal footing.  
Here we will show how a
`superspin' description in terms of a four component order
parameter field may be set up. Three of these will simply
correspond to the Neel vector while the fourth will be the VBS
order parameter (which is a scalar in this case with rectangular
symmetry). A superspin $O(4)$ sigma model (with some weak $O(3)$
anisotropy will be derived. Interestingly this sigma model
contains a topological $\theta$ term (at $\theta = \pi$). We will show how the $O(4)$ sigma model
together with this topological term reproduces the known physics of the coupled spin-$1/2$ chains.

Next we consider spin-$1/2$ quantum magnets on an isotropic two dimensional square lattice.
In interesting recent work Tanaka and Hu\cite{ta-hu} have shown how a `superspin' non-linear sigma
model may be derived to describe the Neel-VBS competition in this system. Such a sigma model again contains a
topological term. Here we revisit their derivation from a slightly different perspective, and show the
equivalence between the resulting sigma model theory and the spinon-gauge descriptions of Refs. \onlinecite{deccp}.
This can be accomplished in full for the case with easy plane anisotropy for the Neel order parameter.

Finally we consider massless $QED_3$ with $N$ two-component Dirac
fermions. For large enough $N$ this is the low energy theory of
the algebraic spin liquid phases shown recently to be stable
quantum phases of $SU(N)$ quantum magnets. The case $N = 4$ is of
direct interest to $SU(2)$ magnets. Here we study $N = 2$ and show
that this is again equivalent to the $O(4)$ sigma model with a
topological term at $\theta = \pi$. The possible duality of $N =
2$ $QED_3$ to the global $O(4)$ model without a topological term
was conjectured recently in Ref. \onlinecite{fvrtx}. Our
derivation provides partial support to this conjecture but reveals
the presence of the topological term.

Based on all these results we
suggest that similar descriptions are possible for massless $QED_3$ at any $N$. If correct this would be a
possibly useful alternate to the spinon-gauge
descriptions that ae currently available.

\section{Algebraic spin liquid in one dimension}
The classic example of a gapless `algebraic' spin liquid is the ground state of the nearest neighbor spin-$1/2$ 
chain. This state has slow power law correlations for both the Neel and VBS correlations. In this brief section we 
provide a lightning overview of its field theoretic description which will be useful for us later. 
The low energy physics of the spin-$1/2$ chain may be described by the following field theory
\begin{eqnarray}
S & = & S_0 + S_{WZW} \\
S_0 & = & \int d^2x \frac{1}{2g} tr(\partial_i U^{\dagger} \partial_i U) \\
S_{WZW} & = & i\Gamma
\end{eqnarray}
Here $U \in SU(2)$. The number $\Gamma$ is defined as follows. The field $U$ defines a map from
$S^2$ to $S^3$. The volume in $S^3$ bounded by the surface traced out by $U$ defines $\Gamma$.
Formally
\begin{equation}
\Gamma = \frac{1}{12\pi} \int d^3y \epsilon_{ijk} tr(U^{-1}\partial_iU U^{-1}\partial_j U
U^{-1}\partial_k U)
\end{equation}
Here the two dimensional space-time $S^2$ is regarded as the boundary of a solid ball B, and $y = (y_1,y_2,y_3)$ are
coordinates of B. The matrix $U$ has been extended to the ball B in such a way that at the boundary it has the
correct value for the two dimensional field theory. We may write $U$ in the form
\begin{equation}
U = \phi_0 + i\vec \phi. \vec \sigma
\end{equation}
with $\phi_0^2 + (\vec \phi)^2 = 1$, {\em i.e} in terms of a four component unit vector. Then the action may
be written
\begin{eqnarray}
S_0 & = & \frac{1}{g} \left(\partial_i \hat\phi)\right)^2 \\
S_{WZW} & = & \frac{i}{6\pi} \int d^3y \epsilon_{ijk} \epsilon_{\alpha\beta\gamma\delta}\phi_\alpha
\partial_i \phi_\beta \partial_j \phi_\gamma \partial_k \phi_\delta
\end{eqnarray}
The Neel and VBS order parameters are simply determined in terms of the unit vector $\hat{\phi}$. We
have $\vec N \sim \vec \phi$ and $D \sim \phi_0$. The WZW field theory has global $SO(4)$ symmetry due to which the Neel vector can be rotated into the 
VBS order parameter. This symmetry which emerges as a property of the low energy fixed point guarantees 
that the Neel and VBS order parameters have the same long distance correlations.
Thus the WZW field theory provides a `superspin' field
theory of the two dominant competing orders of the quantum spin-$1/2$ chain. Note that the topological
WZW term is required to correctly reproduce the known physics of the spin chain in this superspin decription.

The WZW theory must be contrasted with another popular field theory for the spin-$1/2$ chain: 
the $O(3)$ sigma model with a topological term. This representation treats the Neel and VBS orders on unequal footing
even though they eventually have the same low energy behavior. Indeed the VBS order parameter is represented 
rather non-trivially as the topological charge density and its connections to the Neel vector are not at all obvious
in this field theory. 

Finally it is possible to examine the spin-$1/2$ chain in a  slave particle  description of the spins by 
writing the spin operator at site $r$ as 
\begin{equation}
\vec S_r = \frac{f^{\dagger}_{r} \vec \sigma f_r}{2}
\end{equation}
with $f_r$ a spinful fermionic `spinon' field at each lattice site. 
Following standard techniques the 
spin Hamiltonian may be recast as a $U(1)$ (or $SU(2)$) gauge theory of these 
spinons coupled to a fluctuating gauge field - see Refs. \onlinecite{hoso,mrfr}.
A suitable continuum approximation can then be made to obtain a theory of massless Dirac spinons coupled to a 
fluctuating gauge field (massless $QED_2$ in the $U(1)$ case). The gauge fluctuations can then be integrated 
out exactly after bosonization and the resulting theory is equivalent to the WZW theory above. In particular
as emphasized in Ref. \onlinecite{mrfr} the slave spinon is quite distinct from the true spinon known to exist 
in the spectrum of the spin-$1/2$ chain. Thus in this one dimensional algebraic spin
liquid the spinon-gauge description is unnnecessary and a description in terms of the slow competing orders is possible.

\section{Weakly coupled spin chains}
\label{quasi1d}
\subsection{Ordered phases and transitions}
Here we consider a two dimensional system obtained as an infinite array of one
dimensional spin-$1/2$ chains coupled together by short-ranged antiferromagnetic interactions.
The entire system then has rectangular symmetry in two dimensions.
In general such coupling tends to stabilize ordering of one of the slow fluctuating order parameters
of the decoupled chain. For unfrustrated interchain couplings, collinear Neel ordering at $(\pi,\pi)$ is stabilized.
With frustration that suppresses Neel order, columnar VBS ordering at $(\pi, 0)$
often gets stabilized as shown in recent work
by Starykh and Balents\cite{stbal}. The competition between the Neel state and the columnar VBS state
of this rectangular lattice may be understood
in the following terms. Consider the limit of reasonably strong two dimensional coupling. Then
an $D = 2+1$ $O(3)$ non-linear sigma model description of the
Neel vector fluctuations is useful. This model must be supplemented
with appropriate Berry phases. These Berry phases only affect singular hedgehog configurations
of the Neel vector in space-time\cite{hald88}. Due to the Berry
phases these hedgehog insertions have the same symmetry as the VBS order parameter and hence may be
identified with them\cite{ReSaSuN}. With rectangular symmetry it is known that in any putative continuum
description the hedgehogs are doubled (note contrast to square lattice where
they are quadrupled). Neel order is destroyed if the hedgehogs proliferate. Their Berry
phases then induce VBS order. Here the resulting state is two-fold degenerate due to the doubling
of hedgehogs and may be identified with the two-fold degenerate columnar VBS pattern.
The Neel-VBS transition in this case is presumably first order due to the likely relevance of doubled
monopoles at the critical fixed point of the hedgehog-free theory\cite{KM,mv}. 
(However there will be an interesting deconfined
multicritical point).

Now we show how these results may be reproduced in an alternate approach that directly attempts a `superspin' description
of the two competing orders (Neel and VBS) in terms of a four component
field. As already illustrated by the analysis in $ d = 1$, this will require
incorporating an appropriate topological term. Before supplying the derivation
let us first understand the structure of the allowed superspin field theory.
With a four component `order parameter' field, we might attempt a description in terms of
an $O(4)$ non-linear sigma model with some $O(3)$ anisotropy. Ignoring the
anisotropy for the time being, the action for such a theory is
defined by
\begin{equation}
S_0 = \int d^2x d\tau \frac{1}{t}
\left(\partial_{i}\hat{\phi}\right)^2
\end{equation}
where $\hat{\phi}$ is a four-component unit vector. The Neel and VBS order parameters are
related to $\hat{\phi}$ in the same way as in the $d = 1$ case. The order parameter clearly lives in
$S^3$. In three spacetime dimensions (with for instance boundary
conditions that identify space-time with $S^3$) configurations of
the $\hat{\phi}$ field may be classified by an integer `winding
number' corresponding to $\Pi_3(S^3) = Z$. The winding number is
expressed in terms of the $\hat{\phi}$ field through
\begin{equation}
Q = \frac{1}{12\pi^2}\int d^2x d\tau
\epsilon_{ijk}\epsilon_{\alpha\beta\gamma\delta} \phi_\alpha
\partial_i\phi_\beta\partial_j\phi_\gamma\partial_k\phi_\delta
\end{equation}
It is then possible to consider adding a topological Berry
phase term to the action $S_0$ which gives different weights to
the different topological sectors parametrized by $Q$. Thus
consider
\begin{equation}
\label{o4theta}
S = S_0 + i\theta Q
\end{equation}
The physics is clearly periodic under $\theta \rightarrow \theta +
2\pi$. The system is parity and time reversal invariant for
$\theta = 0$ or $\theta = \pi$.

Now we will show that a derivation of such a theory with a
four-component field for the coupled spin chains naturally lead to
the sigma model at $\theta = \pi$. Our derivation will closely
follow Haldane's well-known derivation of the $O(3)$ sigma model
field theory with the $\theta$ term for {\em one dimensional} spin
chains starting from the microscopic path integral for a single
quantum spin. Here we will start instead from the WZW theory for
each spin chain, and then derive a two dimensional continuum field
theory for the field $\hat{\phi}$.

We start from the decoupled chain limit in which each chain is
described by the $SU(2)$ level 1 theory. We then include interchain
couplings for the slow modes of this theory. Consider therefore
the following action for the coupled chains:
\begin{eqnarray}
S & = & S_1 + S_2  \\
S_1 & = & \sum_y S[U(x, \tau,y)]\\
S_2 & = & \int dx d\tau \sum_y [u \vec{\phi}(x, \tau,
y).\vec{\phi}(x, \tau, y+1) \\
& & -v \phi_0(x, \tau,y)\phi_0(x,\tau,y+1)]
\end{eqnarray}
Here $x$ is the continuous coordinate along the chain direction,
$y$ is an integer labelling the different chains, and $U$ is the
$SU(2)$ matrix field which enters the description of each chain.
The action $S[U]$ is the $SU(2)$ level 1 WZW action summed over
all the chains. The term $S_2$ in the action is the interchain
coupling. It has been written in terms of the four component
vector $\hat{\phi}$. The couplings $u,v > 0$. We have assumed there
is an antiferromagnetic coupling of the Neel order parameter
($\vec{\phi}$) between two neighboring chains, and a ferromagnetic
coupling between the VBS orders ($\phi_0$). This is completely
appropriate to describe the competition between $(\pi,\pi)$ Neel
order and $(\pi, 0)$ columnar VBS order in the two dimensional
problem.

It is useful to first consider the simple limit where $u =v$ and
then to perturb around this limit. When $u =v$ the action has
extra global $O(4)$ symmetry that is broken to $O(3) \times Z_2$
when $u \neq v$. At $u =v$, we  can write $S_2$ as
\begin{equation}
S_2 = -\frac{u}{2}\int_{x,\tau}\sum_y tr(U(y+1)U(y))
\end{equation}
This tends to prefer a `staggered' ordering of the $U$
perpendicular to the chains where $U$ for even chains is the
inverse of that for the odd chains. We therefore write
\begin{eqnarray}
\label{contpar}
U(y) & = & g(y)~~~~y =2n \\
& = & g^{\dagger}(y)~~~~y = 2n+1
\end{eqnarray}
and take the continuum limit with $g$ assumed to vary slowly in
both spatial directions. The first term in $S_1$ has a simple
continuum form:
\begin{equation}
S_0 = \frac{1}{t}\int d^2x d\tau tr(\partial_i g^{\dagger}
\partial_i g)
\end{equation}
The continuum form of the WZW term requires some work.
Substituting Eqn. \ref{contpar}, we get
\begin{equation}
\sum_y (-1)^y i \Gamma[g(y)]
\end{equation}
This alternating sum can now be evaluated analogous to Haldane's
calculation in $d = 1$. First write it as
\begin{equation}
\sum_{y = 2n} i (\Gamma[g(y)] - \Gamma[g(y-1)])
\end{equation}
Each term is the difference in volume in $S^3$ bounded by the
surfaces traced out by $g(y)$ and $g(y-1)$. With $g$ smoothly
varying the full sum over $y$ can then be related to the total
volume in $S^3$ swept out by $g(x,y, \tau)$. Specifically we get
\begin{equation}
\int dxdyd\tau \frac{i}{24 \pi} \epsilon_{ijk}tr(g^{-1}\partial_i
g g^{-1}\partial_j g g^{-1}\partial_k g)
\end{equation}
This is precisely the $O(4)$ model at $\theta = \pi$ as may be
seen by writing the $SU(2)$ matrix $g$ in terms of the
four-component $\hat{\phi}$ field.

Taking $u \neq v$ introduces some anisotropy in the model which
lowers the symmetry to $O(3) \times Z_2$. Let us now study this
model in the presence of such weak anisotropy. Consider defect
configurations of the $O(3)$ vector field $\vec \phi$. In three
space-time dimensions these are hedgehogs. In the core of the
hedgehog the magnitude of $\vec \phi$ is suppressed. In the
present model this means that the $\hat{\phi}$ vector points along
the $0$ direction in the core. Clearly we can distinguish two
different kinds of hedgehogs depending on whether $\phi_0$ is
positive or negative in the core. These hedgehogs are very
analogous to meron-vortices in $O(3)$ models with weak easy plane
anisotropy. Therefore we will refer to these as meron-hedgehogs.
Each meron hedgehog may be regarded as one half of a point defect
of the $O(4)$ model with non-zero $Q$. Thus for elementary meron
hedgehogs we have $Q = \pm 1/2$ depending on the sign of $\phi_0$
in the core. The topological $\theta$ term then gives a Berry
phase $e^{\pm i\frac{\pi}{2}}$ associated with each meron
hedgehog.

Let us now study various phases of the anisotropic $O(4)$ model.
When the $O(3)$ vector orders (the Neel phase) the meron hedgehogs
or their Berry phases play very little role in the low energy
physics. Consider now phases where the Neel vector is disordered.
This may be usefully discussed in terms of a three dimensional
Coulomb gas of meron hedgehogs which interact with each other
through $1/r$ interactions and where appropriate Berry phase
factors are associated with the hedgehogs. The action for this
Coulomb gas takes the form
\begin{eqnarray}
S & = & S_c + S_B + S_{int} \\
S_c & = & u\sum_r \left(m_{+r}^2 + m_{-r}^2 \right) \\
S_B & = & i\frac{\pi}{2}\sum_r\left(m_{+r} - m_{-r}\right)\\
S_{int} & = & \sum_{rr'} m_r V(r - r')m_{r'}
\end{eqnarray}
Here $r$ is the site of some three dimensional space-time lattice.
$m_{\pm r}$ are integers corresponding to the number of the two
kind of  $\pm$ hedgehogs at site $r$. The first term is a hedgehog
core action. The second term is the Berry phase. In the last term
$m_r = m_{+r} + m_{-r}$ is the total hedgehog number at any site
$r$. This term describes the interhedgehog three dimensional
Coulomb interaction with $V(r-r') \sim \frac{1}{r-r'}$. As usual
an equivalent sine Gordon theory is readily obtained by decoupling
the interaction with a potential $\chi$ and summing over the
integers $m_{\pm}$. It is easy to see that the resulting action
takes the form
\begin{equation}
\label{sg} S = \int d^3x K \left(\nabla \chi \right)^2 - \sum_n
\lambda_n \cos(2n\chi)
\end{equation}
Here $e^{i\chi}$ creates a strength-$1$ hedgehog. We thus see that
only even strength hedgehogs appear in this sine Gordon theory.
The Berry phases have lead to a doubling of the hedgehogs. The doubled hedgehogs proliferate
at long scales in this paramagnetic phase. This leads to a two-fold degenerate ground state
(the two ground states being distinguished by the expectation value of the single
hedgehog operator). Thus phases that have unbroken $O(3)$ symmetry of the 
anisotropic $O(4)$ model at $\theta = \pi$ necesarily has a two-fold ground state
degeneracy. 

To understand better the nature of these $O(3)$ symmetric phases in the particular microscopic realization
of weakly coupled chains let us study the role of various discrete symmetries. In particular we must 
distinguish between parity transformations $P_x, P_y$ along the chain direction (the $x$-direction) and the one 
perpendicular to the chains. Microscopically the $P_x$ symmetry involves $x \rightarrow -x$ {\em together} with 
$\phi_0 \rightarrow -\phi_0$. For the hedgehogs this then implies $m_+ \rightarrow -m_-, m_- \rightarrow -m_+$, 
$\chi \rightarrow -\chi$. Under $P_y$ we have $y \rightarrow -y, m_{\pm} \rightarrow -m_{\pm}, 
\chi \rightarrow -\chi + \pi$. 

In the smooth ground states of the sine-Gordon model the possible values of $\chi$ depend on the 
${\lambda_n}$. If $\lambda_1$ is the dominant coupling then $\lambda_1 < 0$ prefers $e^{i\chi} = \pm i$
so that $P_x$ is broken while $P_y$ is preserved. We may identify this with the $(\pi,0)$ columnar dimer state. 
On the other hand $\lambda_1 > 0$ prefers $e^{i\chi} = \pm 1$ so that $P_y$ is broken while $P_x$ is preserved. 
This may actually be identified with the $(0, \pi)$ columnar dimer state ({\em i.e} vertical dimers between the chains). 
It is interesting that though we derived the sigma model focussing on the competition between Neel and the 
$(\pi,0)$ VBS state it is still capable of describing the $(0,\pi)$ VBS state as well.

All of this is completely consistent with results known from the earlier direct analysis of the $O(3)$
sigma model with Berry phases appropriate for a two dimensional lattice with rectangular symmetry.
Thus the superspin $O(4)$ formulation with the $\theta$ term correctly captures the Neel-VBS competition in
this system.

Finally we note that spin-$1/2$ magnets on lattices with rectangular symmetry are not expected to have 
trivial featurless paramagnetic phases that also preserve all lattice symmetries. Non-trivial 
paramagnetic phases such as spin liquids with topological order are however possible. This then implies that the 
$O(4)$ model at $\theta = \pi$ in the presence of $O(3) \times Z_2$ anisotropy will also not possess simple
phases which break no symmetry. Any such symmetry unbroken phase must necesarrily also have some 
hidden order (such as topological order).

\section{Deconfined criticality in the two dimensional square lattice}
\label{dccp}
We now turn our attention to spin-$1/2$ quantum magnets on isotropic two dimensional square lattices.
Here apart from the well-known Neel state, paramagnetic VBS states are again possible. In contrast to the
quasi-two dimensional case here the VBS state will have a four fold ground state degeneracy. A superspin description
of the Neel-VBS competition was recently derived by Tanaka and Hu\cite{ta-hu}. Here we first very briefly review their
derivation from a slightly different perspective. In the presence of some easy plane anisotropy for the Neel vector we
explicitly show the equivalence to the spinon-gauge field theory proposed in Ref. \onlinecite{deccp}
for the deconfined critical point between these two ordered phases. The appropriate
spinon-gauge field theory involves gapless
bosonic spinons coupled to a gapless non-compact $U(1)$ gauge field and has been dubbed the non-compact $CP^1$ (NCCP$^1$)
model\cite{mv}.

\subsection{Tanaka-Hu superspin field theory}
Consider a half-filled Hubbard model on a two dimensional square lattice
with $\pi$ flux through each square plaquette described by the Hamiltonian
\begin{equation}
H = -\sum_{<rr'>} t_{rr'}  \left(c^{\dagger}_rc_{r'} + h.c \right) + U \sum_r n_r(n_r - 1)
\end{equation}
Here $c_{r\alpha}$ are spinful electron operators at sites $i$ of a square lattice.
The $t_{rr'}$ are chosen to have $\pi$ flux through each plaquette. A specific choice is
$t_{rr'} = - (-1)^y t$ for horizontal bonds and $t_{rr'} = t$ for vertical bonds. The $U$ term is the
usual onsite Hubbard repulsion. At $U = 0$ the band structure is well-known and consists of four
Fermi points (at $(\pm \frac{\pi}{2}, \pm \frac{\pi}{2})$) in the Brillouin zone of the square lattice.
The electron dispersion is linear
in the vicinity of these points. The low energy physics may be described in terms of a continuum
Dirac theory that focuses on the modes near these Fermi points. For $r = (x, 2y)$ write
\begin{eqnarray}
c_{\vec r} & = & e^{i \vec K_1. \vec r}\psi_{12} + e^{i\vec K_2. \vec r}\psi_{21} \\
c_{\vec r+ \hat{y}} & = & i\left( e^{i \vec K_1. \vec r}\psi_{11} + e^{i\vec K_2. \vec r}\psi_{22} \right)
\end{eqnarray}
with $\vec K_1 = (\frac{\pi}{2}, \frac{\pi}{2})$, and $\vec K_2 = (-\frac{\pi}{2}, \frac{\pi}{2})$.
Then the continuum Hamiltonian reads
\begin{equation}
H \approx \int d^2x \psi^{\dagger} \left( -i \partial_x \tau^z -i \partial_y \tau^x \right)\psi
\end{equation}
Here $\psi = \psi_{a\alpha}$ and each $\psi_{a\alpha}$ is a two-component Dirac spinor. The
Pauli matrices $\vec \tau$ act on the Dirac index. The index $a = 1,2$ labels the node and
$\alpha$ labels the physical spin. The corresponding action (after letting $y \rightarrow -y$) is
\begin{eqnarray}
S & = & \int d^2x d\tau \bar{\psi}\left(-i\tau_y \partial_{\tau} -i \partial_x \tau^x - i \partial_y \tau^z \right)\psi \\
& \equiv & \int d^3x \bar{\psi}\left(-i\tau_i \partial_i \right)\psi
\end{eqnarray}
where we have defined $\bar{\psi} = i\psi^{\dagger}\tau_y$.

At small $U$ the interactions renormalize to zero and a semi-metallic state is stable. As $U$ is increased there
will be a metal-insulator transition where the fermions acquire a charge gap. In the resulting insulator
it is expected that in some parameter ranges a simple Neel state will emerge.
In other parameter ranges a valence bond solid state may well be stabilized. Either insulator may conveniently
be described starting from the Hubbard model in a Hartree-Fock description that builds in a non-zero value for the
appropriate order parameter. To describe the competition between these two kinds of ordered insulators, we will introduce
fields that couple to either order parameter and integrate out the fermions to obtain effective sigma model descriptions.
It is easy to see that the Neel vector coresponds to the fermion bilinear $i\bar{\psi}\vec \sigma \mu_y \psi$
where $\vec \mu$ are Pauli matrices in the node index. The VBS order parameter $(v_x, v_y)$ corresponds to
$v_x = i\bar{\psi}\mu_x \psi, v_y = i\bar{\psi}\mu_z \psi$. To access either insulating phase we
therefore include the terms
\begin{equation}
u_N\left(\bar{\psi}\vec \sigma \mu_y \psi \right)^2 + u_v \left[\left(\bar{\psi}\mu_x \psi\right)^2
+ \left(\bar{\psi}\mu_z \psi \right)^2 \right]
\end{equation}
To begin with we will assume that $u_N = u_v = u$ for simplicity. Later we will add anisotropy between the
Neel and VBS fluctuations. This term may be decoupled with a 5-component Hubbard Stratonovich field
$\vec \phi = (\vec N, \phi_{vy}, \phi_{vx})$. We therefore get
\begin{equation}
S = \int d^3x \bar{\psi}\left(-i\tau_i \partial_i +i \vec \phi .\vec \Gamma \right)\psi +\frac{|\phi|^2}{2u}
\end{equation}
Here $\vec \Gamma$
are a set of five $4 \times 4$ matrices defined through
\begin{eqnarray}
\Gamma_{1,2,3} & = & \sigma_{x,y,z} \mu_y \\
\Gamma_4 & = & \mu_z\\
\Gamma_5 & = & \mu_x
\end{eqnarray}
Note that $[\Gamma_{\alpha}, \Gamma_{\beta}]_+ = 2\delta_{\alpha\beta}$ and
$\Gamma_5 = - \Gamma_1\Gamma_2\Gamma_3\Gamma_4$. Equivalently the phase diagram and universal aspects of the various phases will be preserved by
 restricting the $\phi$ vector to have unit magnitude. We thus consider the action
\begin{equation}
S = \int d^3x \bar{\psi}\left(-i\tau_i \partial_i +i m\hat{\phi} .\vec \Gamma \right)\psi
\end{equation}
The Hartree-Fock aproximation to the insulating phases consists of giving $\hat{\phi}$ some non-zero mean value
which gaps out the fermions. To access these insulators while including fluctuation effects we integrate out the
$\psi$ assuming that $m$ is large. The resulting fermion determinant has been evaluated by 
Abanov and Wiegmann\cite{abwie}
and gives
\begin{equation}
S[\hat{\phi}] = \int d^3x \frac{1}{2g} \left(\partial_i \hat{\phi} \right)^2 - 2\pi i \Gamma[\hat{\phi}]
\end{equation}
This takes the form of a non-linear sigma model for the superspin vector $\hat{\phi}$ that
combines the Neel and VBS order parameters. The crucial feature is the presence of the second term - this is a
Wess-Zumino-Witten term for the 5-component unit vector field $\hat{\phi}$ and is defined as follows. The field
$\hat{\phi}$ defines a map from $S^3$ to $S^4$. The fraction of the total volume of $S^4$ that is bounded by the
hypersurface traced out by $\hat{\phi}$ defines $\Gamma$. Specifically let $\hat{\phi}(x,u)$ be any smooth extension of
$\hat{\phi}(x)$ such that $\hat{\phi}(x,0) = (1,0,0,0,0)$ and $\hat{\phi}(x,1) = \hat{\phi}(x)$. Then
\begin{equation}
\Gamma = \frac{3}{8\pi^2} \int_0^1 du \int d^3x \epsilon_{\alpha\beta\gamma\delta\kappa} \phi_{\alpha}
\partial_x \phi_{\beta} \partial_y \phi_{\gamma} \partial_{\tau} \phi_{\delta} \partial_u \phi_{\kappa}
\end{equation}
Thus the Neel-VBS competition for spin-$1/2$ magnets on an isotropic two dimensional square lattice
is described by this $SO(5)$ superspin non-linear sigma model with the extra topological WZW term.
This sigma model must be supplemented with an anisotropy that reduces the symmetry to $SO(3) \times U(1)$
where the $SO(3)$ corresponds to spin rotation and the $U(1)$ rotates between the two components of the VBS
order parameter. Further microscopically this $U(1)$ symmetry must have further anisotropy
that reduces the symmetry to $Z_4$.

In their original derivation Tanaka and Hu viewed the $\pi$ flux state as a
fermionic spinon mean field theory for Heisenberg spin magnets.
Their subsequent calculations are identical to what we described above. However from that point of view it becomes
important to include gauge fluctuations before establishing a firm connection to the original
quantum magnet\footnote{For the $\pi$ flux state this involves coupling the fermionic spinons to a gapless
$SU(2)$ gauge field}. Here we have sidestepped this issue by considering the Hubbard model on the square lattice with
$\pi$ flux as our starting point. There is no question of introducing extra fluctuating gauge fields from this point of view.
Note that once the system enters the insulating phase the
spin physics at energy scales below the charge gap may be described (in principle) in terms of a short ranged
spin model of spin-$1/2$ moments. Thus the $\pi$-flux Hubbard model as a microscopic starting point
is very convenient to derive an effective theory to describe the Neel-VBS competition in two dimensional quantum magnets.

\subsection{Equivalence to NCCP$^1$ field theory}

We now show the equivalence of the superspin sigma model with the WZW term to the NCCP$^1$ model 
in Ref. \onlinecite{deccp} for the deconfined critical point between the Neel and VBS states. We will do
this in the specific case where there is large easy plane anisotropy on the Neel vector so that it points
primarily in the $xy$ plane in spin space. To handle this case we write
\begin{equation}
\hat{\phi} \approx (0, \hat{\Pi})
\end{equation}
with $\hat{\Pi} = (\vec N_{\perp}, \phi_{vy}, \phi_{vx})$ a four component unit vector. Here $\vec N_{\perp}$ is
a two component vector which denotes the direction of the Neel vector in spin space. With this restriction the
Wess-Zumino term in the action can be evaluated straightforwardly and simply becomes
\begin{equation}
S_{WZW} = -i\pi Q
\end{equation}
where $Q$ is the by now familiar index describing the winding of
the unit four vector field $\hat{\Pi}$ in three space-time
dimensions. Thus we appear to be back to the $O(4)$ sigma model at
$\theta = \pi$ but now with $O(2) \times O(2)$ anisotropy for the
$\hat{\Pi}$ field\footnote{We will ignore the $Z_4$ anisotropy on
the VBS order parameter for the present discussion. It can be
reinstated without any major changes if necessary and in any case
has been argued to be irrelevant at the deconfined critical fixed
point.}

The topological term again plays a crucial role. To expose its role let us warm up by considering some
simple configurations of the $\hat{\Pi}$ field. It will be useful to view it as a two component complex
vector $z = (z_1, z_2)$ with $z_{1,2}$ complex numbers satisfying $|z_1|^2 + |z_2|^2 = 1$. The topological index
$Q$ is readily writen in terms of derivatives of $z$. An equivalent form is obtained by considering the ``vector
potential"
\begin{equation}
a_i = -i z^{\dagger}\partial_i z .
\end{equation}
Some algebra shows that
\begin{equation}
Q = \frac{1}{2\pi} \int d^3x \epsilon_{ijk}a_i \partial_j a_k
\end{equation}
As a topological index $Q$ is invariant under smooth continuous deformations of $z(x)$. As a particular example a smooth
gauge transformation $z \rightarrow e^{i\theta(x)} z$ changes $a_i$ to $a_i + \partial_i \theta$. The integral
for $Q$ above is clearly invariant under such a gauge transformation.

Consider first configurations where $z_1 = |z_1| e^{i\theta_1(x)}$ with $\theta_1$ a smooth single valued function of $x$,
and $z_2$ arbitrary.
This means that that there are no vortices anywhere in the complex field $z_1$. Then the gauge transformation
$z \rightarrow e^{-i\theta_1(x)}z$ makes $z_1$ real everywhere. It is clear then that for such configurations $Q = 0$
(as with $z_1$ real one component of $\hat{\Pi}$ vanishes everywhere). The same result obviously also holds for
configurations where there are no vortices anywhere in $z_2$ but $z_1$ is arbitrary. Thus $Q$ is non-zero only if
there are vortices in both $z_1$ and $z_2$.

\begin{figure}
\includegraphics[width=4.8cm]{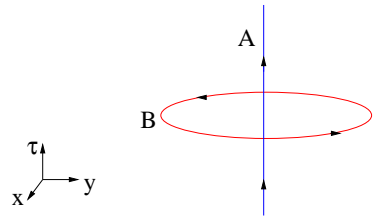}
\caption{Configuration with $Q = 1$ showing the location of the vortices in $z_1$ and $z_2$. The curve A is the 
vortex in $z_2$ and B is the vortex in $z_1$.} \label{vrtxlnk}
\end{figure}

To construct a configuration with non-zero $Q$ consider the map
\begin{equation}
\hat{\Pi}(r, \theta, \phi) = \left(cos(\alpha(r)), sin(\alpha(r)\hat{e_r}\right)
\end{equation}
with $(r, \theta, \rho)$ the spherical coordinates for space-time and $\alpha(r)$ some smooth function of $r$
satisfying $\alpha(0) = 0$ and $\alpha(\infty) = \pi$. Here $\hat{e_r}$ is a radial unit vector
defined by $(cos\theta, sin\theta cos\phi, sin\theta sin\phi)$ (the first component is along the
$\tau$ direction). This configuration clearly has $Q = 1$ and is a simple generalization of the familiar
$O(3)$ skyrmion in two dimensions. Let us identify the location of the vortex lines of $z_{1,2}$ in this
configuration. When $\theta = 0$, $z = (e^{i\alpha}, 0)$. Similarly when $\theta = \pi$, $z = (e^{-i\alpha}, 0)$.
Thus along the line $A$ (see Fig. \ref{vrtxlnk}) 
which runs along the $\tau$ axis at the spatial origin $z = (e^{i\theta_1}, 0)$ with
$\theta_1$ varying from $-\pi$ to $\pi$. Similarly at $r$ such that $\alpha(r) = \frac{\pi}{2}$ and
$\theta = \frac{\pi}{2}$ ({\em i.e} along the curve $B$) we have $z = (0, e^{i\phi})$. Thus we may identify $A$
with a vortex line in $z_2$ and $B$ with a vortex line in $z_1$. In this configuration these two
oriented vortex lines have a non-trivial linking number $1$.

These considerations illustrate a general mathematical result which
we will henceforth use: The topological index $Q$ may be
identified with the linking number between oriented vortex loops
in $z_1$ and those in $z_2$. For the $\theta = \pi$ model this
implies that whenever a $z_1$-vortex moves fully around a
$z_2$-vortex (or vice versa) there is a phase factor of $\pi$. In
other words the $z_1$ and $z_2$ vortices are mutually non-local
and see each other as sources of $\pi$-flux. 

It is now
straightforward to see the equivalence to the easy-plane NCCP$^1$
action. Indeed the vortices in $z_1$ are just the $2\pi$ vortices in the easy plane Neel 
order parameter. The $\pi$ phase shift that these vortices 
acquire when taken around the vortices in $z_2$ suggests that 
the latter may be thought of as spin-$1/2$ spinons. This is indeed consistent with 
the expected structure of the vortices in the 
VBS order parameter\cite{levsent}. To explicitly 
demonstrate the equivalence to the NCCP$^1$ action let us go to a dual representation in terms of the
vortices in $z_1$ and $z_2$. A lattice action which incorporates
the non-trivial mutual statistics of these two vortices is readily
written down:
\begin{eqnarray}
S & = & S_{t1} + S_{t2} + S_a + S_{CS} \\
S_{t1} & = & -t\sum_{<ij>} \sigma_{ij}cos(\vec \nabla \phi_1
-\vec a_1)\\
S_{t2} & = & -t \sum_{<IJ>}\mu_{IJ} cos(\vec \nabla \phi_2 - \vec a_2)\\
S_a & = & K\sum_P \left(\vec \nabla \times \vec a_1 \right)^2 +
\left(\vec \nabla \times \vec a_2 \right)^2 \\
S_{CS} & = & i\sum_{<IJ>}\frac{\pi}{4}(1 - \mu_{IJ})(1 - \prod_P
\sigma)
\end{eqnarray}
Here $i,j,...$ are the sites of a cubic space-time lattice and
$I,J, ...$ the sites of the corresponding dual lattice. The fields
$\phi_{1,2}$ are the phases of the vortices in $z_1$ and $z_2$
respectively. These vortices are coupled minimally to their
respective non-compact $U(1)$ gauge fields $a_{1,2}$. The
$\sigma_{ij}, \mu_{IJ}$ are $Z_2$ gauge variables that take values
$\pm 1$. The last term is an Ising mutual Chern-Simmons
term\cite{z2long} that incorporates the mutual $\pi$ statistics of
the two vortices. This will be made explicit below. The symbol
$\prod_P \sigma$ represents a product of $\sigma_{ij}$ over the
four links of the spacetime plaquette pierced by the dual link
$<IJ>$.

To check that the $S_{CS}$ term correctly implements the mutual statistics of the 
two vortices it is convenient to go to a Villain representation. For instance for the 
$2$-vortex the hopping term becomes
\begin{equation}
S_{t2}  \rightarrow  \sum_{<IJ>} u(\vec j_{2})^2 +i\vec j_{2}.(\vec
\nabla \phi_2 - \vec a_2 +\frac{\pi}{2}(1- \mu_{IJ})
\end{equation}
A sum over the integer vortex currents $\vec j_2$ is implied. The
integration over $\phi_2$ as usual implies the current
conservation condition
\begin{equation}
\label{vcons}
 \vec \nabla . \vec j_2 = 0
\end{equation}
The sum over the Ising variable $\mu_{IJ}$ may also be performed
and leads to the constraint
\begin{equation}
\label{mutpi}
 (-1)^{j_2} = \prod_P \sigma
\end{equation}
Thus the presence of a $2$-vortex is seen as $\pi$ flux by the
$1$-vortex.  Thus the term $S_{CS}$ precisely implements the
mutual $\pi$ statistics of the two vortices. 

In the Neel phase the vortex $e^{i\phi_1}$ is gapped while the 
$2$-vortex is condensed (with the reverse being true in the VBS phase). At a putative second order transition between the 
two, we might expect that both vortices stay critical.  
A standard duality transformation
on one of the two vortex degrees of freedom enables us to demonstrate that 
this action is indeed in the easy plane NCCP$^1$ universality class near this transition. 
This is elaborated in Appendix \ref{dual}. 

Note that the easy plane deconfined critical fixed point also has $O(2) \times O(2)$ symmetry\cite{deccp} 
consistent with the 
superspin field theory discussed in this section. One of these $O(2)$ symmetries is realized simply as a 
uniform phase rotation of both the matter fields $e^{i\theta_{\pm}}$ of Appendix \ref{dual}. 
The other $O(2)$ is realized non-trivial as a topological symmetry associated with the conservation of the 
photon.

\section{Sigma model description of massless $QED_3$}
In this section we will consider massless $QED_3$ in $D = 2 + 1$ dimensions.
This consists of $N$ two-component Dirac fermions $\psi$ coupled to a fluctuating non-compact
$U(1)$ gauge field $a$
with the action
\begin{equation}
S = \int d^3x \bar{\psi}\left(\tau_i(-i \partial_i - a_i) \right)\psi +
\frac{1}{2e^2}\left(\epsilon_{ijk} \partial_j a_k \right)^2
\end{equation}
The $\tau_i$ are Pauli matrices. This action has a global $SU(N)$ symmetry asociated with unitary rotations of $\psi$
and a hidden global $U(1)$ symmery associated with gauge flux conservation.
For $N = 4$ this is the low energy description of the dRVB algebraic spin liquid of $SU(2)$ invariant
spin-$1/2$ models that has been much discussed recently.
Neither the Dirac fermions nor the gauge photon are well-defined quasiparticle excitations
of the system when its low energy properties are controlled by a scale invariant fixed point (as happens
generically for large enough $N$ and at special multicritical points for small $N$).
For any $N \geq 2$ this scale invariant theory has slow power law
correlations for a set of gauge invariant fermion bilinears that transform as adjoints under the $SU(N)$.
Can we dispense with the gauge description completely in favor of some sort of sigma model decription
in terms of these slow fluctuations? As discussed in previous sections 
this is certainly possible in various other analogous problems but required 
inclusion of a topological term in the sigma model action.
Here we study the specific case $N = 2$ in $D = 2 +1$ and show that a sigma model description
is again possible provided topological terms are included.

For $N = 2$ the fermions transform with spin-$1/2$ under the global $SU(2)$ symmetry. It is expected that
the gauge invariant $SU(2)$ vector $\bar{\psi}\sigma\psi$ will have slow correlations, and perhaps might
even generically order. To expose its effects let us add the following term to the action
\begin{equation}
S_{4} = \int d^3x u\left(\bar{\psi}\sigma\psi \right)^2
\end{equation}
with $u >0$. This term may be decoupled with a fluctuating $\vec N$ field to get
 a term of the form
 \begin{equation}
 \frac{(\vec N)^2}{2u} +i \vec N. \bar{\psi}\sigma\psi
 \end{equation}
 Equivalently the phase diagram and universal aspects of the various phases will be preserved by
 restricting the $\vec N$ vector to have unit magnitude. We thus consider the action
 \begin{eqnarray}
 S & = & S_f + S_a \\
 S_f & = & \int d^3x \bar{\psi}\left(\tau_i(-i \partial_i - a_i) + im \hat{n}.\vec \sigma \right)\psi \\
 S_a & = & \frac{1}{2e^2}\left(\epsilon_{ijk} \partial_j a_k \right)^2
 \end{eqnarray}
 with $m > 0$.
 The fields $\psi, a, \hat{n}$ are all to be integrated over. We begin by first doing the
quadratic integral over the fermion fields. We assume space-time to be $S^3$.
The resulting fermion determinant has been calculated
(in a $1/m$ expansion) by Abanov and Wiegmann\cite{abwie} and leads to the following remarkable result:
\begin{eqnarray}
S & = & S_0 + S_j + S_H + S_a \\
S_0 & = & \int d^3x \frac{1}{g} \left(\partial_i \hat{n} \right)^2 \\
S_j & = & \int d^3x iJ_ia_i \\
S_H & = & -i\pi H[\hat n]
\end{eqnarray}
Here $g \sim 1/m$ in $S_0$. The $J_i$ is the topological current density of the $\hat{n}$ field, {\em i.e}
it is the skyrmion current density. Formally
\begin{equation}
J_i = \frac{1}{8\pi} \epsilon_{ijk} \hat{n}. \partial_j \hat{n} \times \partial_k \hat{n}
\end{equation}
The last term of the action involves an interesting topological invariant $H$ corresponding to
$\Pi_3(S^2) = Z$ and is known as the Hopf term. $H$ is an integer that distinguishes different space-time
configurations of the $\hat{n}$ field. $H$ is conveniently written in terms of a $CP^1$ spinor $z$
associated with the field $\hat{n}$ (defined through
$\hat{n} = z^{\dagger}\vec \sigma z$). From $z$ form the $SU(2)$ matrix $U = [z -i\sigma^yz^*]$. Then
\begin{equation}
H = \frac{1}{24 \pi^2}\int d^3x \epsilon_{ijk} tr(U^{-1}\partial_iU U^{-1}\partial_j U
U^{-1}\partial_k U)
\end{equation}
In the absence of the coupling to the gauge field, the Hopf term changes the spin and statistics of the
skyrmion.
Note that the gauge field $a_i$ couples to the skyrmion current density of the
$\hat{n}$ vector. Therefore at long scales the skyrmion density is pinned at zero. Thus
the resulting model is the three dimensional $O(3)$ vector model where skyrmion configurations have been
suppressed. This leads to enlargement of the degrees of freedom to a four component field and the
symmetry to $O(4)$. To see this most simply, rewrite the $O(3)$ sigma model above in the $CP^1$ representation.
We have
\begin{eqnarray}
S_0 & = & \int d^3x \frac{1}{g} |\left(\partial_i -i A_i \right) z|^2 \\
S_j & = & \int d^3x \frac{i}{2\pi}\epsilon_{ijk} a_i \partial_j A_k
\end{eqnarray}
We may now integrate over the gauge field $a$. This is conveniently done by choosing a gauge where
$A_i$ is transverse ({\em i.e} $\vec \nabla. \vec A = 0$). We get
\begin{eqnarray}
S & = & S_z + S_H \\
S_z & = & \int d^3x \frac{1}{g} |\left(\partial_i -i A_i \right) z|^2 + \frac{e^2}{8\pi^2} A_i^2
\end{eqnarray}
Thus the $CP^1$ gauge field $A_i$ has been rendered massive due to the coupling to the fluctuating gauge field
$a$. At long distances and low energies we may thus drop the gauge field altogether.
(Strictly speaking we
should just integrate out the gauge field to generate a term that is quartic in the $z$ fields and
involves two derivatives -
this term is expected to be irrelevant at the critical point of the resulting theory. A similar result 
was established in Ref. \cite{o4} in the absence of the topological term). The action $S_z$ then becomes
\begin{equation}
S_z  = \int d^3x \frac{1}{g} |\partial_i z|^2
\end{equation}
which has $O(4)$ symmetry as can be made manifest by rewriting in terms of the real and
imaginary parts of the two components of $z$. What happens of the Hopf term? Clearly $S_H$ is invariant
both under independent left and right multiplications by constant $SU(2)$ matrices. Thus it too is $O(4)$ invariant.
A simple calculation shows that the Hopf invariant $H$ is exactly equal to $Q$ where $Q$ is
the topological index characterizing
configurations of an $O(4)$ unit vector field in three dimensions. Thus the action reduces to the $O(4)$ model at
$\theta = \pi$.

Note that in the $O(4)$ symmetry broken phase there are three gapless linear dispersing modes. In the gauge theory
description this corresponds to broken chiral symmetry with $<\hat{n}> \neq 0$. Here again there are three gapless
linear dispersing modes - two are spin waves in $\hat{n}$ while the third is simply the gapless photon.

A somewhat similar possible duality between massless $N = 2$ $QED_3$ and the usual critical $O(4)$ model was conjectured 
recently in Ref. \onlinecite{fvrtx} using very different arguments. Our derivation shows that such a duality indeed exists 
but necessarily includes the topological term. 
Does the topological term make any difference to the properties of the model? We turn to this question in the following subsection. 

\subsection{Isotropic $O(4)$ model with a $\theta$ term}
What may we say about the properties of the {\em isotropic} ({\em i.e} $O(4)$ symmetric) model at $\theta = \pi$
from the analysis in this paper? First we expect that a phase with broken $O(4)$ symmetry is stable for weak coupling. The
$\theta$ term presumably plays very little role at low energies in this phase as it only affects
topological configurations that cost large energy. At strong coupling paramagnetic phases where $O(4)$ symmetry
is preserved are presumably possible. The arguments in previous sections show that a trivial featureless paramagnet
cannot exist. To see this
first assume that such a paramagnetic phase can indeed exist. Then turn on a small anisotropy 
that breaks the $O(4)$ symmetry to $O(3) \times Z_2$. Such an anisotropy
will have no effect on the ground state of such a trivial paramagnet - but the arguments of Section \ref{quasi1d}
shows that
such trivial paramagnetic ground states do not exist in the anisotropic model. Thus such states are forbidden
for the isotropic model as well. What are the possibilities then? Gapped phases can exist
if the ground state has topological order.
Gapless paramagnetic states are also not precluded by these arguments.

Similar and more interesting conclusions may also be drawn by considering weak $O(2) \times O(2)$ anisotropy (instead of $O(3)
\times Z_2$). First note that with this anisotropy the model has an extra $Z_2$ symmetry that interchanges the 
two $O(2)$ symmetries. It is clear that this is exactly equivalent to the 
Motrunich-Vishwanath self-duality symmetry\cite{mv} of the equivalent easy plane NCCP$^1$ model. We can now discuss the 
phase diagram of the $\theta = \pi$ $O(4)$ model with weak $O(2) \times O(2)$ anisotropy in terms of the known phases and phase transitions 
of models with the same field content as the easy plane NCCP$^1$ model. First consider the $O(4)$ ordered phase. 
With the $O(2) \times O(2)$ anisotropy this becomes the location of a first order `spin flop' transition between 
phases that separately break either of the two distinct $O(2)$ symmetries. What about phases that preserve $O(2) \times
O(2)$ symmetry? Here there are two possibilities. First the self-dual 
second order deconfined critical line of the easy plane NCCP$^1$ appears as a paramagnetic {\em phase} of the sigma model. 
Thus this is a concrete example of the possibility of a gapless strong coupling paramagnetic phase at $\theta = \pi$
in the $O(4)$ model (albeit with some $O(2) \times O(2)$ anisotropy). It is concievable that the self-dual critical 
manifold of the easy plane NCCP$^1$ admits a special multicrtitical point with higher $O(4)$ symmetry. The 
existence of such a fixed point would imply the possibility of a gapless paramagnetic phase at $\theta = \pi$ with full 
$O(4)$ symmetry. The other possibility that preserves full $O(2) \times O(2)$ symmetry is a gapped $Z_2$ 
topologically ordered paramagnet which coresponds to the $Z_2$ spin liquid allowed in the 
lattice easy plane spin-$1/2$ antiferromagnet. This phase should persist with full $O(4)$ symmetry as well. 

In view of the above it seems clear that the transitions out of the ordered phases with $O(2) \times O(2)$ 
anisotropy would be in a different universality class 
from those in a model without the $\theta$ term. With full $O(4)$ symmetry
it thus seems rather likely that at $\theta = \pi$ the transition is in a different universaity class from that at 
$\theta = 0$.

\section{Discussion and prospects}
In this paper we have provided a number of examples illustrating how the physics of two dimensional spin-$1/2$ 
quantum magnets can in principle be described in `superspin' sigma models that describe the slow competing orders. 
As such these are as close that one might get to a Landau-Ginzburg description of these competing orders. 
An important feature of these sigma models is the presence of topological terms which reflect the 
underlying quantum nature of the spins. From the conventional point of view it is 
these topological terms that complicate a direct application of naive Landau-Ginzburg-Wilson(LGW) thinking to 
quantum magnetism. Thus they are at the root of the failure of the LGW paradigm in describing various quantum 
phases and phase transitions. 

The sigma model formulations provide a potential alternate to the gauge theoretic descriptions that have been 
used thus far to describe the non-LGW physics. However as things stand we know even less about how to handle 
the effects of topological terms than we know about gauge theories. So it is at present not clear how useful 
the sigma model will be. 

Finally our results raise the possibility of such a sigma model description for stable algebraic spin liquids 
in two dimensions (of which the $dRVB$ state popular in cuprate physics may be a possible example). These also have slow 
power law corelations in a number of physical observables. The results on $N = 2$ massless $QED_3$ provide some 
positive hints. Stronger evidence is the sigma model description of the deconfined Neel-VBS critical point. 
These may be thought of as a special kind of algebraic spin liquid that has one relevant perturbation. 
In the gauge theory description they have gapless bosonic matter fields coupled to a non-compact $U(1)$ gauge field.
As these theories seem to have sigma model descriptions perhaps their fermionic cousins do as well. Perhaps such 
descriptions might even be useful!

\section*{Acknowledgements}
We thank M. Freedman, M. Hermele, O. Motrunich, and Diptiman Sen for useful discussions. 
This work was initiated at the Aspen Center for Physics Summer Progarm on ``Gauge Theories 
in Condensed Matter Physics". 
TS also
acknowledges funding from the NEC Corporation, the Alfred P. Sloan
Foundation, and an award from the The Research Corporation.
MPAF was supported by the National Science Foundation
through grant
DMR-0210790.

\begin{appendix}
\section{Duality of the $\theta = \pi$ $O(2) \times O(2)$ model to NCCP$^1$}
\label{dual}
Proceeding as usual
we solve the current conservation condition Eqn.  \ref{vcons} by
writing
\begin{equation}
\vec j_2 = \vec \nabla \times \vec A
\end{equation}
with $\vec A$ an integer living on the links $<ij>$. 
We have for the action
\begin{eqnarray}
S & = & S_{t1} + S_A + S_{aA}+ S_a \\
S_A & = & u\sum_P (\vec \nabla \times \vec A)^2  \\
S_{aA} & = & \sum_{IJ} \vec a_2. \vec \nabla \times \vec A
\end{eqnarray}
with $S_{t1}$ and $S_a$ as before. This must be supplemented with the mutual statistics 
condition
\begin{equation}
\prod_P \sigma_{ij} (-1)^{A_{ij}} = 1
\end{equation}
To handle ths we write 
\begin{equation}
A_{ij} = 2A'_{ij} + s_{ij}
\end{equation}
with $s = 0,1$ and $A'$ an integer. Then we have
\begin{equation}
\prod_P \sigma_{ij} (-1)^{s_{ij}} = 1
\end{equation}
This is solved by
\begin{equation}
\sigma_{ij} (-1)^{s_{ij}} = \alpha_i \alpha_j
\end{equation}
with $\alpha_i = \pm 1$. 
The integer condition on $A'$ can be implemented (softly) with a term 
\begin{eqnarray}
-t_ccos\left(2\pi A'_{ij}\right) & = & -t_ccos\left(\pi(A_{ij} - s_{ij})\right) \\
& = & -t_c\sigma_{ij}\alpha_i \alpha_j cos\left(\pi A_{ij} \right)
\end{eqnarray}
We now separate out the longitudinal part of $\vec A$:
\begin{equation}
\vec A = \vec A^T + \frac{1}{\pi}\vec \nabla \theta_c
\end{equation}
Collecting all the pieces of the action together and integrating out the gauge field $\vec a_2$
we see that $\vec A^T$ is massive. We will therefore simply drop it from now on. We are then left 
with
\begin{eqnarray}
S & = & S_{t1} + S_{tc} + S_{a1} \\
S_{tc} & = & -t_c \sum_{ij} \sigma_{ij} cos\left(\vec \nabla \theta_c \right) \\
S_{a1} & = & K\sum_P \left(\vec \nabla \times \vec a_1 \right)^2
\end{eqnarray}
In writing the term $S_{tc}$ we have absorbed the $\alpha_i$ into the $\theta_{ci}$. The sum over 
the $Z_2$ gauge field $\sigma_{ij}$ may now be straightforwardly performed. It generates a number of terms
of which the most important have the structure
\begin{equation}
-\kappa \left(cos(\vec \nabla \theta_+ - a_1) + cos(\vec \nabla \theta_- + a_1) \right)
\end{equation}
where $\theta_{\pm} = \theta_c  \pm \phi_1$. Together with the term $S_{a1}$ this is precisely the action for 
the easy plane NCCP$^1$ theory. 

\end{appendix}


\begin{thebibliography}{99}

\bibitem{deccp}T. Senthil, A. Vishwanath, L.
Balents, S. Sachdev, and Matthew P.A. Fisher, Science 303, 1490
(2004); T. Senthil, L. Balents, S. Sachdev, A. Vishwanath, and M.
P. A. Fisher, Phys. Rev. B 70, 144407 (2004).

\bibitem{stable-u1}M. Hermele, T.
Senthil, M. P. A. Fisher, P. A. Lee, N. Nagaosa, and X.-G. Wen,
Phys. Rev. B 70, 214437 (2004).

\bibitem{rantwen}  W. Rantner and X.-G. Wen, Phys. Rev. B 66, 144501 (2002).

\bibitem{su4}M. Hermele, T. Senthil, and M. P. A. Fisher,
Phys. Rev. B 72, 104404 (2005).

\bibitem{hald83}F.D.M. Haldane, Phys. Lett. 93A, 464 (1983); Phys. Rev. Lett. 50, 1153 (1983);
 J. Appl. Phys. 57, 3359 (1985).  

\bibitem{1dwzw} I. Affleck and F. D. M. Haldane, Phys. Rev. B 36, 5291 (1987). 

\bibitem{hoso}Y. Hosotani, J. Phys. A 30, L757, (1997); D. H. Kim and P.A. Lee, Annals Phys. 272 (1999) 130-164

\bibitem{mrfr}C. M. Mudry and E. Fradkin, Phys. Rev. B 50, 11409 (1994).

\bibitem{hald88}F. D. M. Haldane, Phys. Rev. Lett. 61, 1029 (1988).

\bibitem{ReSaSuN} N. Read and S. Sachdev, Phys. Rev. Lett. 62, 1694
  (1989); Phys. Rev. B 42, 4568 (1990).

\bibitem{ta-hu}A. Tanaka and X. Hu,
Phys. Rev. Lett. 95, 036402 (2005)

\bibitem{fvrtx}J. Alicea, O. I. Motrunich, M. Hermele, and M. P. A. Fisher,
Phys. Rev. B 72, 064407 (2005)


\bibitem{stbal}O. A. Starykh and L. Balents,
Phys. Rev. Lett. 93, 127202 (2004)

\bibitem{KM}M. Kamal and G.
  Murthy, Phys. Rev. Lett. 71, 1911-1914 (1993)


\bibitem{mv}O. I. Motrunich and A. Vishwanath
Phys. Rev. B 70, 075104 (2004)

\bibitem{abwie} A. G. Abanov and P. B. Wiegmann, Nucl. Phys. B 570, 685 (2000).

\bibitem{levsent} M. Levin and T. Senthil, Phys. Rev. B 70, 220403 (2004).


\bibitem{z2long} T. Senthil and M. P. A. Fisher,
Phys. Rev. B 62, 7850-7881 (2000)

\bibitem{o4} A. V. Chubukov, S. Sachdev and T. Senthil, Nucl. Phys. B 426, 601 (1994)








\end{thebibliography}
\end{document}